\documentclass[12pt]{article}
\usepackage{graphicx,amsfonts,amsmath,amssymb,mathrsfs}

\title{Canonical Typicality}

\author{Sheldon Goldstein$^*$\footnote{Departments of Mathematics and
Physics, Hill Center, Rutgers, The State University of New Jersey, 110
Frelinghuysen Road, Piscataway, NJ 08854-8019, USA}\footnote{E-mail:
oldstein@math.rutgers.edu}\ ,  Joel  L. Lebowitz$^*$\footnote{E-mail:  
lebowitz@math.rutgers.edu}\ ,\\ Roderich
Tumulka\footnote{Mathematisches Institut, Eberhard-Karls-Universit\"at,
Auf der Morgenstelle 10, 72076 T\"ubingen, Germany.  E-mail:
tumulka@everest.mathematik.uni-tuebingen.de}\ , and Nino
Zangh\`\i\footnote{Dipartimento di Fisica dell'Universit\`a di Genova
and INFN sezione di Genova, Via Dodecaneso 33, 16146 Genova, Italy.
E-mail: zanghi@ge.infn.it} } \date{December 27, 2005}

 \newcommand{\Hilbert}{\mathscr{H}}
 \newcommand{\tr}{\mathrm{tr}}
\renewcommand{\Im}{\mathrm{Im}}
   \renewcommand{\Re}{\mathrm{Re}}

\renewcommand{\sp}[2]{\langle #1|#2 \rangle}
\newcommand{\I}{{[E,E+\delta]}} 

 \newcommand{\sys}{{(S)}}
\newcommand{\hb}{{(B)}} \newcommand{\rhored}{\rho}
\newcommand{\rhocan}{\rho_\beta}

\begin{document}

\maketitle \begin{abstract}

It is well known that a system, $S$, weakly coupled to a heat bath, $B$, is
described by the canonical ensemble when the composite, $S+B$, is described
by the microcanonical ensemble corresponding to a suitable energy shell.
This is true both for classical distributions on the phase space and for
quantum density matrices.  Here we show that a much stronger statement
holds for quantum systems.  Even if the state of the composite corresponds
to a single wave function rather than a mixture, the reduced density matrix
of the system is canonical, for the overwhelming majority of wave functions
in the subspace corresponding to the energy interval encompassed by the
microcanonical ensemble. This clarifies, expands and justifies remarks made
by Schr\"odinger in 1952.

\medskip

 \end{abstract}

A quantum system in thermal equilibrium at inverse temperature $\beta$
is described by the canonical density matrix \begin{equation}
\label{rhocandef} \rho _\beta  = \frac{1}{Z} \exp(-\beta H^{(S)})
\end{equation} where $H^{(S)}$ is the system Hamiltonian and $ Z =\tr\,
\exp(-\beta H^{(S)})$. The usual justification for \eqref{rhocandef} is
that it is the reduced density matrix of the system when  it is weakly
coupled to a heat bath, and the composite system is described by the
microcanonical density matrix at a suitable total energy $E$.

More explicitly, one assumes that it is permissible to neglect the
relatively small interaction between the system and the bath, so that
the total Hamiltonian of the composite $S+B$ is given by
\begin{equation} H = H^{(S+B)}= H^{(S)} + H^{(B)}\,. \end{equation} The
composite $S+B$ is then assumed to be represented by a microcanonical
ensemble in some energy interval $[E, E+\delta]$, where $\delta$  is
small on the macroscopic scale,  $\delta \ll E$,  but large enough for
the interval to contain very many eigenvalues. The corresponding
microcanonical density matrix  is \begin{equation}\label{microcanrho}
\rho_{E,\delta} = (\dim \Hilbert_{\I})^{-1} P_{\Hilbert_{\I}}
\end{equation} with $P_{\Hilbert_{\I}} $ the projection to
$\Hilbert_{\I}$, the spectral subspace for $H$ associated with energies
in the interval $\I$ in the Hilbert space $\Hilbert = \Hilbert^{(S+B)}
=\Hilbert^{(S)} \otimes \Hilbert^{(B)}$. One readily proves (see below)
that in the thermodynamic limit,  when the size, i.e., the number of
components $N$ of the heat bath, goes to infinity while $E/N \to e$,
the reduced density matrix  of the system $S$
\begin{equation}\label{rdm} \rho^{(S)} = \mbox{tr}^{(B)} \rho_{E,
\delta} \end{equation} is equal to $\rho_\beta$.  Here $\tr^{(B)}$
denotes the partial trace over $\Hilbert^\hb$ and $\beta = \beta(e)$.

In this note we show how this result can be substantially strengthened: we
prove that, in the thermodynamic limit, {\em the reduced density matrices
of the overwhelming majority of the wave functions of $S+B$ are
canonical}. We call this statement \emph{canonical typicality}. As a
consequence of canonical typicality, it follows that the canonical ensemble
is even more inevitable in quantum mechanics---arising as it does there
without the invocation of any genuine randomness---than it is classically.
Results in this direction were first obtained by Schr\"odinger
\cite{schrbook}, and later by Gemmer and Mahler \cite{gm}; related results
have been obtained by Tasaki \cite{Tasaki1}; see later.

Typicality in quantum mechanics, as well as in classical mechanics,
involves a probability distribution on the possible microstates of the
system, the distribution in terms of which ``overwhelming majority'' is
to be understood.  In classical mechanics these microstates are points
in the appropriate phase space and the distribution is then a measure
on this phase space. To define typicality for quantum systems we shall
take the microstates to be wave functions, i.e., points on the unit
sphere of $\Hilbert$ (up to a phase). Even with this identification it
may not be clear which distribution is appropriate for the (composite)
system  described by the density matrix $ \rho_{E, \delta}$.  Here we
take that to be the probability distribution proposed long ago by
Schr\"odinger \cite{schr,schrbook} and Bloch \cite{Bloch}: it is the
(normalized) uniform (surface area) measure $u_{E,\delta}$ on the unit
sphere in the subspace $\Hilbert_{[E,E+\delta]}$, i.e., the uniform
probability distribution over all normalized wave functions $\Psi$ with
energies in $\I$. If we expand $\Psi$ in terms of energy eigenfunctions
   $|E_\alpha\rangle$  of $H$, $ \Psi = \sum  c_\alpha |E_\alpha 
\rangle$,
where the sum is restricted to levels with energies in the interval
$\I$, then $u_{E,\delta}$ corresponds to the uniform distribution on
the surface of the sphere $ \sum |c_\alpha|^2 =1$. This measure was
shown in \cite{Bloch} to  be, in a well defined sense, the most
appropriate distribution corresponding to  the microcanonical density
matrix,  i.e., such that \begin{equation}\label{me} \rho_{E,\delta} =
\int  u_{E,\delta} (d\Psi) |\Psi\rangle\langle \Psi |\, .
\end{equation}

Let $\rhored^\Psi$ denote the reduced density matrix of the system,
given that $S+B$ is in a pure state $\Psi \in  \Hilbert_{\I}$,
\begin{equation}\label{rhoreddef} \rhored^\Psi = \tr^\hb \, |\Psi
\rangle \langle \Psi |\,. \end{equation} We may then ask: for which
wave functions $\Psi$ is $\rhored^\Psi$ (approximately) of the
canonical form \eqref{rhocandef}.  We make the standard assumption that
both $H^\sys$ and $H^\hb$ have pure point spectrum and are bounded {} 
from
below.  In $\Hilbert^\sys$ ($\Hilbert^\hb$) we choose an eigenbasis of
$H^\sys$ ($H^\hb$), denoted $|E^\sys_1 \rangle, |E^\sys_2 \rangle,
\ldots{}$ ($|E^\hb_1 \rangle, |E^\hb_2 \rangle, \ldots{}$), with
corresponding eigenvalues $E^\sys_1 \leq E^\sys_2 \leq \ldots{}$
($E^\hb_1 \leq E^\hb_2 \leq \ldots{}$).

First of all we note the following: saying that for the majority of
$\Psi$'s, $\rho^\Psi$ is close to $\rhocan$ is equivalent to saying
that if $\Psi$ is \emph{randomly} chosen with distribution
$u_{E,\delta}$ then with overwhelming \emph{probability}, $\rho^\Psi$
is close to $\rhocan$. From now on we will thus regard $\Psi$ as a
(Hilbert-space-valued) random variable.

We begin by recalling the standard derivation of the canonical ensemble
{}from the microcanonical. One has for the reduced density matrix
$\rho^{(S)}$ of the system that \begin{equation} \rho^{(S)}  =
(\mbox{dim}\,\Hilbert_\I)^{-1}\sum_i \mbox{dim} (\Hilbert^{(B)}_i )
|E^\sys_i \rangle \langle E^\sys_i|\,, \label {star}\end{equation}
where \begin{equation} \Hilbert^{(B)}_i = \Hilbert^{(B)}_{[E
-E^\sys_i,\, E -E^\sys_i +\delta]} \end{equation} is the spectral
subspace for $H^{(B)}$ associated with energies in the interval $
[E -E^\sys_i, E -E^\sys_i +\delta]$. It is then more or less clear, and
can be rigorously proven under suitable conditions, that when the bath
is sufficiently large $\rho^{(S)}\approx\rho_\beta $, with $\beta =
dS(E)/dE$ where $S(E)$ is the bath's entropy. This follows {}from the
basic fact that $S(E)\approx \log \mbox{dim} (\Hilbert^{(B)}_{[E,
E+\delta]} ) $,  so that
\begin{equation}
\mbox{dim} (\Hilbert^{(B)}_i )\approx e^{
S(E-E^\sys_i)} \approx e^{ S(E)-\beta E^\sys_i} \sim e^{-\beta E^ 
\sys_i}\,.
\end{equation}
More precisely, one proves that $\rho^{(S)}\to \rho_\beta$ in the
thermodynamic limit with $\beta = ds(e)/de$, where $s(e) = \lim
[S(E)/N]$ and $e=\lim[ E/N ] $.

Thus to demonstrate canonical typicality it suffices to establish that
(\ref{star}) holds, at least approximately, when $\rho^{(S)}$ is
replaced by $\rho^\Psi$ for typical $\Psi\in \Hilbert_\I$. A key step
of our argument is to note that the uniformly distributed random vector
$\Psi$ can always be regarded as arising by normalization
\begin{equation} \Psi = \frac{\Phi}{\| \Phi \|} \end{equation} {}from a
Gaussian random vector $\Phi \in \Hilbert_\I$ with mean zero and
covariance given by the identity operator on $\Hilbert_\I$. This  means
that in the decomposition \begin{equation}\label{aaaa} \Phi = \sum_{i}
\sum_{j} C_{ij} |E^\sys_i \rangle |E^\hb_j \rangle\,, \end{equation}
the real and imaginary parts, $\Re \, C_{ij}$ and $\Im \, C_{ij}$, of
the coefficients are independent real Gaussian random variables with
mean zero and variance $1/2$ for those $i$ and $j$ for which $E^\sys_i
+ E^\hb_j \in \I$ (and $C_{ij} =0$ otherwise).  We obtain {}from
(\ref{aaaa}) that \begin{equation} \Phi = \sum_i |E^\sys_i \rangle
|\Phi_i \rangle \end{equation} with \begin{equation}\label{uuu} |\Phi_i
\rangle = \sum_{j\in I_i} C_{ij}|E^\hb_j \rangle\,, \end{equation}
where $I_i$ is the set of bath levels  $j$ such that $E^\hb_j \in [
E-E^\sys_i, E -E^\sys_i +\delta]$, whence the reduced density matrix
(\ref{rhoreddef}) is of the form \begin{equation}\label{rhoredu}
\rhored^\Psi = \frac{1}{\|\Phi\|^2} \tr^\hb \, |\Phi \rangle \langle
\Phi | = \frac{1}{\|\Phi\|^2} \sum_{i,i'} \sp{ \Phi_i}{\Phi_{i'}} \,
|E^\sys_i \rangle \langle E^\sys_{i'} |\,. \end{equation}

Now, if  $\delta$ were so small   that the system's energy spacings
$\Delta E^\sys_i =E^\sys_{i+1} - E^\sys_i$ are all greater than $\delta
$, then the relevant energy intervals $I_i$ for the heat bath would be
pairwise disjoint and the $\Phi_i$ pairwise orthogonal,
\begin{equation}\label{ortho} \sp{ \Phi_i}{\Phi_{i'}} = \delta_{ii'}
\|\Phi_i\|^2 \,.  \end{equation} We argue now that equation (\ref{ortho})
will continue to hold, at least approximately, even without the above
assumption on $\Delta E^\sys_i$. This is so because, when $I_i$ and
$I_{i'}$ have significant overlap, the contributions to $\Phi_{i}$ and
$\Phi_{i'}$ corresponding to the sum over the terms in (\ref{uuu})
belonging to both $I_i$ and $I_{i'}$ will typically be approximately
orthogonal. To see this note that these sums form two independent random
vectors, with uniformly distributed directions, in a high-dimensional
space.  As such they are, with probability approaching unity, nearly
orthogonal since the expected value of the absolute square of the dot
product of two independent random unit vectors in an $n$-dimensional space
is, by symmetry, $1/n$.

 From the representation (\ref{uuu}) we have that \begin{equation}
\|\Phi_i\|^2 = \sum_j |C_{ij}|^2\,, \end{equation} so that
$\|\Phi_i\|^2 $ is the sum of the $N_i=\dim ( \Hilbert^\hb_i)$
independent, identically distributed random variables $|C_{ij}|^2$ with
mean 1.  It thus follows, by the law of large numbers, that typically
\begin{equation} \|\Phi_i\|^2 \approx \dim ( \Hilbert^\hb_i)\,.
\end{equation} We thus have for the reduced density matrix
(\ref{rhoreddef}), using (\ref{rhoredu}), that typically
\begin{equation} \rho^\Psi   \approx (\dim \Hilbert_\I)^{-1}\sum_i
\mbox{dim} (\Hilbert^{(B)}_i ) |E^\sys_i \rangle \langle E^\sys_i|\,,
\label {BB} \end{equation} which is what we needed to show.

Concerning (\ref{BB}), we remark that it follows merely {}from the fact
that the reduced density matrix $\rho^\Psi$ does not depend upon $\Psi$
for typical $\Psi \in \Hilbert_{E,\delta}$, that whenever the reduced
microcanonical density matrix $\rho^{(S)}\approx \rho_\beta$, the same
is true for $\rho^\Psi$ for typical $\Psi$: applying the partial trace
$\mbox{tr}^{(B)}$ to (\ref{me}), one obtains that $ \rho^{(S)} = \int
u_{E,\delta} (d\Psi) \rho^\Psi \approx \rho^\Psi$, for typical $\Psi$.
\bigskip

Some essential parts of the argument we have presented here have
already been described by Schr\"odinger in an appendix, written in
1952, to his book on \textit{Statistical Thermodynamics}
\cite{schrbook}. We note that Schr\"odinger in  \cite{schrbook} made
the assumption that ``in a big system  \ldots{} the amplitude- 
squares  \ldots{} are
on the average [in time] equal for \ldots{} eigenfunctions
belonging to the same [energy] eigenvalue \ldots{}.''    He uses this
assumption for the combined system  $S+B$  to derive   ``exactly the
same canonical distribution between the amplitude-squares, as is in the
customary treatment said to indicate the probability of the system
being on this or that energy level.''  However, Schr\"odinger neither  
connects the assumption
with typicality, nor his conclusion with the reduced density matrix
$\rho^\Psi$, which he does not even mention. His concern is rather   
with showing
that one need not regard a system in thermal equilibrium   as being in
a  energy eigenstate. As he states in his preface, ``To ascribe
to every system always one of its sharp energy values is an
indefensible attitude.''

Tasaki \cite{Tasaki1} has studied, as we do here, the reduced density
matrix of a system coupled to a heat bath when $S+B$ is described by a wave
function $\Psi$.  He shows that for a special form of the coupling
Hamiltonian the long-time average of $\rho^{\Psi(t)}$ is canonical. He then
argues in a heuristic way that also for typical large times,
$\rho^{\Psi(t)} \approx \rhocan$. (In this argument, there is a hidden
typicality assumption on the initial wave function $\Psi(0)$, namely that
all of the energy expansion coefficients are small, which is true of most
wave functions.) His argument does not yield our stronger statement that
$\rho^{\Psi(t)} \approx \rhocan$ even at $t=0$ for typical wave
functions. Tasaki also includes some examples that he studied rigorously,
concerning energy eigenstates of the composite or states that are initially
product states $\Psi^\sys \otimes \Psi^\hb$ with $\Psi^\sys$ an eigenstate
of $H^\sys$.

More recently, Gemmer and Mahler \cite{gm} have established canonical
typicality under the assumption of very large degeneracy for energy
eigenstates by computing appropriate Hilbert space volumes.

\bigskip

\noindent \textit{Acknowledgments.}  We appreciate the hospitality  
that some of
us have enjoyed, while working on this paper, at the Institut des  
Hautes \'Etudes Scientifiques
(Bures-sur-Yvette, France), the Mathematics
Department of Rutgers University, and the Dipartimento di Fisica of
Universit\`a di Genova. The work of S.~Goldstein was supported in  
part by NSF Grant
DMS-0504504, and that of J.~Lebowitz by NSF Grant DMR 01-279-26 and  
AFOSR
Grant AF 49620-01-1-0154. The work of R.~Tumulka was supported by  
INFN and
by the European Commission through its 6th Framework Programme
``Structuring the European Research Area'' and the contract
Nr. RITA-CT-2004-505493 for the provision of Transnational Access
implemented as Specific Support Action. The work of N.~Zangh\`\i\ was
supported by INFN.

\bigskip

\noindent \textit{Note added:} In a very recent preprint, Popescu, Short,
and Winter  have established canonical typicality under great
generality by invoking Levy's Lemma  \cite{PSW05}.

\end{document}